\documentclass[fleqn,12pt,twoside]{article}
\usepackage{espcrc1}
\usepackage{graphicx}
\newcommand{\AmS}{{\protect\the\textfont2
  A\kern-.1667em\lower.5ex\hbox{M}\kern-.125emS}}
\newcommand{\beq}{\begin{equation}}
\newcommand{\eeq} [1] {\label{#1} \end{equation}}

\title{Multilevel Monte Carlo method for simulations of fluids}

\author{Achi Brandt,  
 Valery Ilyin\address{Department of Computer Science and Applied 
Mathematics \\
Weizmann Institute of Science, Rehovot 76100, Israel}}

\begin{document}
\maketitle

\begin{abstract}
Monte Carlo methods play important part in  modern statistical
 physics. 
The application of these methods suffer from two main difficulties.The first 
is caused by the relatively small number of particles that can participate  
in any numerical calculation. This means that scales larger than or comparable 
to the one that can be simulated  are not taken into account. The second 
difficulty is  the locality of the conventional Monte Carlo algorithms which 
leads to very ( sometimes unreasonably) long  equilibration times. These 
obstacles can be eliminated in the framework of the multilevel Monte Carlo 
method described here. The basic approach is to describe the system at 
increasingly coarser levels  defined on increasingly large domains, 
and  transfer  information back and forth between the levels in order to 
obtain a selfconsistent result. The method is illustrated for a test case 
of  one-dimensional fluids
\end{abstract}

\section{INTRODUCTION}
   The modern theory of classical liquids is based on the statement that
the macroscopic characteristics of a many-particle system can be obtained
by averaging over microscopic configurations, with the probability proportional
to the Gibbs distribution function. Every microscopic configuration is
defined by assigning specific locations to the particles,  and
the main computational problem of  statistical physics consists in the  
difficulty of averaging over the enormous space of possible configurations. 
In order to estimate the value  of this average the Monte Carlo 
technique for the canonical ensemble  was proposed \cite{metr}. 

   However, the straightforward application of this technique is 
restricted to a small volume of the system under consideration, because one 
can consider only a relatively small number  of particles in any numerical 
calculation. In order to minimize the surface effect on the bulk values which 
are to be calculated,  artificial periodic boundary conditions are supposed 
\cite{metr} similar to the case of an infinite crystal \cite{montr}.

This means that at  scales comparable with the periodicity cell size the 
fluctuations of the particle number are cut off. In order to avoid this
difficulty several approaches have been suggested using the grand canonical 
ensemble  \cite{adams},\cite{yao},\cite{karl}. The main trait of these 
algorithms is the generation of density fluctuations in the basic cell by 
adding and deleting  particles in accordance with the value of the chemical 
potential. 

   The Monte Carlo simulations both in canonical and grand canonical ensembles 
are very local, moving, e.g., one particle at a time. The operation of 
 adding/deleting a particle is also local and needs to be followed by a local 
equilibration. This leads to very slow changes of large-scale features, 
such as averages and various types of clusters (regions of (anti)aligned 
dipoles, crystallized segments, etc.).The larger the scale the slower the 
change and longer (per particle!) is the Monte Carlo process required to 
produce new independent features. Since many independent features are needed 
for calculating accurate averages, and since very-large-scale features need 
to be sampled, especially in the vicinity of phase transitions, the 
computations often become extremely expensive, sometimes even losing 
practical ergodicity. In recent years, a number of novel 
Monte Carlo algorithms ensuring ergodicity were proposed \cite{mlt}, 
\cite{binder}, \cite{pan}. In order to avoid the slowness of the ordinary 
Monte Carlo simulation, steps of more collective nature are used. 
Nevertheless, the simulation is performed in the periodicity cell and the 
estimation of bulk values may only be done  by  extrapolation \cite{binder1}. 

An approach which allows to simultaneously overcome slowness and finite 
size effects of the conventional Monte Carlo method consists of a multilevel 
view of the system, realized by multilevel algorithms \cite{brandt11},
\cite{brandt12}. The multilevel algorithms construct a sequence of 
descriptions of the system under consideration at increasingly coarser 
levels and transfer information back and forth between the levels in order 
to obtain a selfconsistent result. The efficiency of multilevel methods 
in solving problems of  statistical physics has been shown on  examples with
sufficiently simple systems \cite{brandt4}. It follows from
these results that the effect of slowing down can be eliminated. Moreover, 
due to the possibility to simulate large volumes  at coarse levels,
the {\it volume factor} (the proportionality of the computational cost to
the volume being simulated) can  be suppressed as well. This means that
the scale of treating a system is not restricted by the size of the 
basic cell. The elimination of both the slowing down and the volume factor 
allows one to investigate in the framework of the multilevel method long 
range phenomena such as phase transition. This possibility has been shown 
on  two-dimensional examples of variable-coefficient Gaussian models 
\cite{brandt4} and the Ising model 
\cite{brandt2}. 

The successful application of the multilevel methods to  lattice 
systems excites interest in adapting  them to more complicated cases. 
The present paper treats multilevel algorithms to the investigation of 
fluids. Numerical results of  simulations for one dimensional models 
are presented. These models reproduce common properties of fluids and allow 
comparisons to exact solutions.  

\section{MULTILEVEL MONTE CARLO APPROACH}
\subsection{Conventional Monte Carlo Method}
The Monte Carlo method  in the statistical theory of liquids  
is used to evaluate 
numerically the average $\overline{A}$ of any functional $A$, defined by:
\beq
\overline{A}=\int\limits_{\Omega} A(\underline{X})\cdot w(\underline{X})
\cdot d\underline{X}\approx \frac{1}{m}\sum\limits^{m}_{i=1} 
A(\underline{X_{i}})
\eeq{c1}
where $w(\underline{X})$ is the probability density of the state 
$\underline{X}$ in the configuration space $\Omega$, and {\it the nodes} 
$\underline{X_{i}}$ are generated by a  random walk in $\Omega$ that 
satisfies detailed balance.

The simplest definition of the  probability to walk from $\underline{X}$ 
to $\underline{X^{\prime}}$ 
in detailed balance  is given by :
\beq
\omega(\underline{X}\to \underline{X^{\prime}})=\min\left[ 1,
 \frac {w(\underline{X^{\prime}})}{w(\underline{X})}\right]\hspace{2mm},
\eeq{c2}
provided that the probability $P(\underline{X^{\prime}}\hspace{.5mm} ?|
\hspace{.5mm} \underline{X})$
for having chosen the state $\underline{X^{\prime}}$ as {\it the candidate} 
to replace the current state $X$ is symmetric, i.e., $P(\underline{X^{\prime}}
\hspace{.5mm}?|\hspace{.5mm}\underline{X})=P(\underline{X}\hspace{.5mm}?|
\hspace{.5mm}\underline{X^{\prime}})$.

The kind of probability density given by Gibbs  in the canonical ensemble 
for simple liquids is:
\beq
w(\underline{X})=\it{const}\cdot \exp 
(-\frac{U(\vec{r}_{1},\vec{r}_{2},\ldots,\vec{r}_{N})}{k_{B}\cdot T})
\eeq{c3}
where $\vec{r}_{i}$ is the location of the $i$-th particle 
($i=1,\ldots ,N$), $k_{B}$ is the Boltzmann constant, T is the 
temperature and $U$ is the potential energy. It is usually assumed to have 
the form:
\beq
U=\sum\limits_{i<j}\phi(\mid \vec{r}_{i}-\vec{r}_{j} \mid )
\eeq{c4}
where $ \phi(\mid \vec{r}_{i}-\vec{r}_{j} \mid   )$ corresponds to the
energy of a two-body interaction. 

In any numerical calculation it is possible to consider only relatively 
small number of particles (extremely small in comparison with Avogadro 
number). In order to minimize the surface effect in a small simulation 
volume,  periodic boundary conditions are supposed \cite{metr}. 
This means that the space is filled by particles which are located at points:
\beq
\vec{R}_{i,\vec{l}}=\vec{r}_{i}+\vec{l}\cdot L,\hbox{   $1\leq i\leq N$}
\eeq{c8}
where $\vec{l}$ is a vector whose components are integer, particles 
at points $\vec{r}_{i}$ are located inside the basic cell of the linear size 
$L$, and $N$ is the particle number in the periodicity cell.

It follows from the periodicity condition that the real system is replaced 
by a super-lattice with the same value of the particle number density 
$\rho=N / L^{D}$ (where $D$ being the dimension of the space) in 
each cell.

   The motion of particles is continuous and in any configuration the 
particles  have arbitrary locations. The transition between states  is made by 
{\it the shift of one particle at a time} \cite{metr} by a small amount via 
the following displacement:
\beq
\vec{r}_{i}^{\hspace{1.5mm}new}=\vec{r}_{i}^{\hspace{1.5mm}old}+\delta_{max}
\cdot (\vec{1}-2\cdot\vec{\xi})
\eeq{c5}
where $\vec{r}_{i}^{\hspace{1.5mm}old}$ and $\vec{r}_{i}^{\hspace{1.5mm}new}$ 
are old and new locations of the $i$-th particle, $\delta_{max}$ is the 
maximum possible displacement of the particle along a coordinate axes, 
$\vec{1}$ is the vector all of whose components are $1$, and $\vec{\xi}$ is 
a vector whose components are random numbers distributed uniformly on the 
interval $[0,1]$.

For shifting the particle with the number $i$, say, one can see  from 
(\ref{c2}) that it is enough to use, instead of the Gibbs function (\ref{c3}) 
with the energy  (\ref{c4}), the {\it conditional} probability defined by:
\beq
P(\vec{r}_{i}\mid{\bf R_{i}})=\it{const}\cdot\exp(-u_{i}({\bf R_{i}})/k_{B}T)
\eeq{c6}
The last relation gives us the probability to find the particle with the
number $i$ at the position $\vec{r}_{i}$ when the locations of all other
particles, defined by the set
${\bf R_{i}}=\{ \vec{r}_{1},\ldots ,\vec{r}_{i-1},\vec{r}_{i+1},
\ldots ,\vec{r}_{N} \} $, are fixed. 
The one-particle energy $u_{i}({\bf R_{i}})$ in (\ref{c6}) is defined by:
\beq
u_{i}({\bf R_{i}})=\sum\limits_{j,(j\neq i)}\phi(\mid \vec{r}_{i}-
\vec{r}_{j}\mid) 
\eeq{c7}
It follows from (\ref{c5}) that the conventional Monte Carlo simulation 
is the local process. Therefore one can expect the high speed convergence 
only in the case of systems with short-range correlations.

\subsection{Coarse variables}
A slowing-down is inherent not only in the conventional Monte Carlo 
algorithm, it is a common problem for all {\it local} processes (e.g. Gauss-
Seidel relaxation for discretizied partial differential equations). The 
solution to this problem lies in introducing system changes of more 
collective nature.  In the case of partial differential equations fast 
convergence of solutions had been attained by multigrid algorithms 
\cite{brandt1}. These algorithms are looking for solutions on a sequence of 
lattices with increasingly larger meshsize (coarser scales) by combining 
local processing at each scale with various inter-scale (inter-lattice) 
interactions. A similar technique can be applied to the simulation of liquids.

A possible way to introduce a coarse description of  liquid consists in 
the discretization of space. The periodicity cell is divided into 
$M$ disjoint parts (e.g. cubes) $V_{i}^{1}$ of equal volume with linear size
$h_{1}$, $1\leq i\leq M$ (each $V_{i}^{1}$ being associated with a gridpoint $i$ 
of the first coarse-level lattice).  Configurations of the finest (particle) 
level are mapped to the first coarse level by the operation of 
{\it coarsening}, this operation creates the coarse-level variable set.
For example, at any instant the corresponding coarse-level variable can be 
defined by coarsening the particle number:
\beq
n_{i}^{1}=\hbox{   Number of particles in  } V_{i}^{1}
\eeq{cv1}
with $\sum\limits_{i=1}^{M}n_{i}=N$, where $N$ is the total number of 
particles in the periodicity cell.
The set $\{ n_{i}^{1}\}$  defines the current configuration on the first 
coarse-level: instead of particle locations the occupation numbers at 
gridpoints are used.

Generally, the  aim of coarsening is the creation of configurations 
represented by 
collective variables which describe collective particle motions  at 
different scales. The variable at each lattice point at each coarse level 
is defined as a {\it local spatial average} (an average or actually a sum 
over a certain neighborhood of the lattice point) of similar variables at the 
next finer level. The total value of each such variable  is well defined for 
each configuration of particles given at the finest level.

The extension of the coarsening  operation (\ref{cv1}) to coarser levels 
leads to the following definition of the coarse-variable at the level $k$:
\beq
n_{j}^{k}=\sum\limits_{V_{i}^{k-1}\subset V_{j}^{k}}n_{i}^{k-1},
\hbox{    $k>1$}
\eeq{cv2} 
for each volume element $V_{j}^{k}$ of level $k$, assuming it to be a union 
of volume elements of the level $k-1$. The coarsening can be repeated till 
the coarsest level, whose choice depends on the scale of the phenomena one 
wants to compute.

The coarser level simulations are performed in periodicity cells of larger 
sizes. In this way finite size effects are suppressed and it is possible to reach the macroscopic description of the system. No slowdown should occur 
provided that the {\it coarsening ratio} (the ratio between a coarse meshsize 
and the next finer meshsize), as well as the average number of of original 
particles per mesh volume of the finest lattice, are suitably low. The typical 
meshsize ratio is 2, typical number of particles per finest lattice mesh 
is between 2 and 10 (being usually larger at a higher dimension). More 
aggressive coarsening ratios would require much longer simulations to 
produce information for coarser levels.

There are many possible ways to choose the set of coarse variables. A general 
criterion for the quality of  this set  is the speed of equilibration of a 
{\it compatible Monte Carlo} (CMC). By this we mean a Monte Carlo process on 
the fine level which is restricted to the subset of fine-level configurations 
whose local spatial averages coincide with a {\it fixed} coarse-level
configuration. A {\it fast} CMC equilibration implies that up to local 
processing all equilibrium configurations are fully determined by their 
coarse-level representations (their local spatial averages), which is the 
main desired property of coarsening.

The compatible Monte Carlo equilibration speed can be tested at the fine 
level, in the following way:  after local thermalization of the ordinary 
Monte Carlo process, for a given configuration in the equilibrium the 
corresponding coarse variables are calculated by (\ref{cv1}) or (\ref{cv2}).
Then an initial configuration in the original variables is 
created in accordance with the coarse variables set. A following ensemble of  
compatible Monte Carlo processes yields an accurate  estimate of the rate of 
approach to the equilibrium. The interpolation from the first coarse level 
configuration to the finest (particle) level needs a special consideration 
owing to the mapping of variables  defined on a grid into the continuous 
space of particle coordinates.
\subsection{CP Tables and Multilevel Algorithm}
The main idea of the multilevel approach is to equilibrate on each level 
only modes with short (comparable with the meshsize) wave lengths. Long 
wave modes with slow convergence at a given level are equilibrated 
at coarser levels where their wave lengths are comparable with the meshsize. 
As a result, the multilevel process leads to fast equilibration of all 
modes.

In the framework of this multilevel Monte Carlo algorithm, only a {\it local} 
process is performed at each level, defined in terms of the corresponding 
variables. In order to calculate transition probabilities (\ref{c2})  
{\it conditional probabilities} similar to (\ref{c6}) should be derived 
for  each coarse level. Such conditional probabilities are expressed in the 
form of a {\it Conditional Probability (CP) table}, which in principle 
tabulates numerically the probability distribution of any pair of neighboring 
coarse-level variables  given the values of all others. Of course, not 
{\it all} other variables should in practice be taken into account: only the 
immediate neighborhood  of gridpoint pair under test counts, due to the 
{\it near locality} property of the conditional probability (cf. also 
discussion of near locality in \cite{brandt2}).

For example, in terms of variables (\ref{cv1}), (\ref{cv2}) defined at 
gridpoints a conditional probability tables $P_{k}(n^{k}_{i},n^{k}_{j}\mid
s^{1}_{i},\ldots ,s^{l}_{i})$ can be constructed from a given sequence 
of configurations in equilibrium on the next finer level $k-1$. These 
tables give us the dependence of the probability to find $n^{k}_{i}$ 
and $n^{k}_{j}$ particles at the two neighboring gridpoints $i$ and $j$ 
on $l$ values in their neighborhood by:

\beq
s^{m}_{ij}=\sum\limits_{\hbox{$q\in neighborhood\hspace{1mm} 
of\hspace{1mm} i\hspace{1mm}and\hspace{1mm} j$}} \alpha^{m}_{q}\cdot n^{k}_{q},
\hspace{2mm} 1\leq m \leq l
\eeq{eq23}
where $\alpha^{m}_{q}$ are  preassigned, suitably chosen coefficients, 
with $\alpha_{i}^{m}=\alpha_{j}^{m}=0$ for all $m$.

For example, in one dimension $j=i+1$ and a possible choice, for $l=1$, 
$\alpha^{1}_{i-1}=\alpha^{1}_{i+2}=1$, otherwise $\alpha^{m}_{q}=0$.

In the coarse level Monte Carlo run, each trial move consists of  particle 
exchange between two neighboring gridpoints, i.e. $n_{i}^{k}\to 
n^{k \prime}_{i}=n_{i}^{k}+\Delta n,\hspace{2mm} n_{j}^{k}\to 
n^{k \prime}_{j}=n_{j}^{k}-\Delta n$.In accordance with (\ref{c2}), the 
acceptance probability for this move is:
\beq
\omega (\underline{X}\to \underline{X}^{\prime})=
\min\left[ 1,
\frac{ P_{k}( n_{i}^{k \prime},n_{j}^{k \prime}\mid s^{1}_{i},\ldots ,
s^{l}_{i})}
{ P_{k}( n_{i}^{k},n_{j}^{k}\mid s^{1}_{i},\ldots ,s^{l}_{i}) } \right]
\eeq{eq24}

The CP tables for any coarse level $k$ are calculated by gathering appropriate
statistics during Monte Carlo simulations {\it at the next finer level} $k-1$. 
Because of the near-locality property, no {\it global} equilibration is 
needed; local equilibration is enough to provide the correct CP values for 
any neighborhood for which enough cases have appeared in the simulation. 
Thus, the fine-level simulation can be done in a relatively small periodicity 
cell.

However, since the fine-level canonical ensemble simulations use only 
a {\it small} periodicity cell, many types of neighborhoods that would
be typical at  some parts of a {\it large} volume (e.g., typical
at parts with average densities different than that used in the periodicity 
cell) will not show up or will be too rare to have sufficiently accurate
statistics. Hence, simulations at some coarse level may run into a situation
in which the CP table being used has flags indicating that values one wants to
extract from it start to have poor accuracy. In such a situation, a temporary
{\it local} return to finer levels should be made, to accumulate more 
statistics, relevant for the new local conditions.

To return from a coarse level to the next finer level one needs first to
{\it interpolate}, i.e., to produce the fine level configurations
represented by the current coarse level configuration, with correct relative
probabilities. The interpolation is performed by  CMC sweeps at the fine 
level (few sweeps are enough, due to the fast CMC equilibration).
This fast equilibration also implies that the interpolation can be done just 
over a restricted {\it subdomain}, serving as a {\it window}:  In the window
interior good equilibrium is reached. Additional passes can then be 
made of {\it ordinary} (not compatible) MC, to accumulate in the interior of 
the window the desired additional CP  statistics, while keeping the window 
boundary frozen (i.e., compatible). The window can then be coarsened (by 
the local spatial averaging) and returned to the coarse level, where 
simulations can now resume with the improved CP table. 

Iterating back and forth between increasingly coarser levels and window
processing at finer levels whenever missing CP statistics is encountered, one
can quickly converge the required CP tables at all levels of the system,
with only relatively small computational domains employed at each level.
The size of those domains needs only  be several times larger than the
size of the neighborhoods being used (with a truncation error that then
decreases exponentially with  that size). However, larger domains 
are better, since they provide sampling of a richer set of neighborhoods 
(diminishing the need for returning later to accumulate more statistics), 
and since the total amount of work at each level depends anyway only on the 
desired amount of statistics, not on the size of the computational domain.
\section{NUMERICAL TESTS}
Thermodynamic properties of a one-dimensional fluid can be described exactly 
under the assumption of the nearest neighbor finite range interparticle 
interaction \cite{taka},\cite{frank}. 
One can cite as an example of the short-range potential the truncated 
Lennard-Jones potential:
\beq
\phi_{LJ}(r)=\Biggl \{
\begin{array}{ll}
4\cdot\epsilon\cdot\biggl[\biggl( \frac{\sigma}{r}\biggr)^{12}
-\biggl( \frac{\sigma}{r}\biggr)^{6}\biggr], & \mbox{for $r\leq r_{cut}$} \\
0,            & \mbox{for $r> r_{cut}$}
\end{array}
\eeq{lj2} 
where $\sigma$ is related to the diameter of atoms, $\epsilon$ is the 
measure of interparticle interaction and $r_{cut}$ is the cut-off 
distance.

At high temperatures the attractive part of the potential can be neglected 
and a good approximation for the interparticle interaction is the hard 
rods model:
\beq
e^{-\frac{\phi_{H}(r)}{k_{B}T}}=
\biggl \{
\begin{array}{ll}
0 & \mbox{,   $r\leq d$} \\
1 & \mbox{,   $r> d$}
\end{array}
\eeq{tl4}
where $d$ is the diameter of every particle.

In order to test the multilevel algorithm it was applied to the simulation 
of one-dimensional fluids. First of all, the speed of the compatible Monte 
Carlo equilibration was tested in the case of the interparticle interaction 
(\ref{lj2}) with $_{cut}=2.5\cdot\sigma$. After locally equilibrating an 
ordinary Monte Carlo process, a sequence of equilibrium configurations was 
picked out. For each configuration the coarse-level variable set is defined 
by (\ref{cv1}) and an initial configuration for compatible Monte Carlo 
simulation is constructed  in accordance with the frozen equilibrium coarse 
level structure.

For the Lennard-Jones fluid a convenient quantity measuring the 
approach to thermal equilibrium is the energy of the system.
However, this quantity is subject to fluctuations during a single run of 
the equilibration. In order to smooth out  simulation results a number of 
compatible Monte Carlo simulations are performed using the same initial 
configuration but different sequences of random numbers. The relaxation 
function is defined by the averaging over these CMC runs:
\beq
R_{U_{0}}(\tau)=\frac{<U(\tau)>-<U(\tau_{f})>}{U_{0}-<U(\tau_{f})>}
\eeq{t2}
where $U(\tau)$ is the energy of the system after sweep $\tau$, subscript 
$f$ means final configurations, and $U_{0}=U(0)$.

The  ensemble average relaxation function (\ref{t2}) is used to
estimate the relaxation time:
\beq
\tau_{eq}(U_{0})=\int\limits_{0}^{\tau_{f}}R_{U_{0}}(\tau)\cdot d\tau
\eeq{t3} 
Here the relaxation time depends on the initial configuration of CMC runs, 
therefore additional averaging of the relaxation time (\ref{t3}) is done 
over all the equilibrium configurations which were chosen during the ordinary 
Monte Carlo process as the initial configuration set for CMC:
\beq
\tau_{eq}=<\tau_{eq}(U_{0})>_{U_{0}}
\eeq{tt3}  

Definitions (\ref{t3}) and (\ref{tt3}) lead to the exact result in the case 
of a single classical relaxation behavior. Nevertheless, even if the 
relaxation behavior is polydispersive, the integral (\ref{t3}) still gives an 
unambiguous measure of the relaxation rate \cite{t.binder}.

\begin{figure}[h]
\centering
\includegraphics[width=4in]{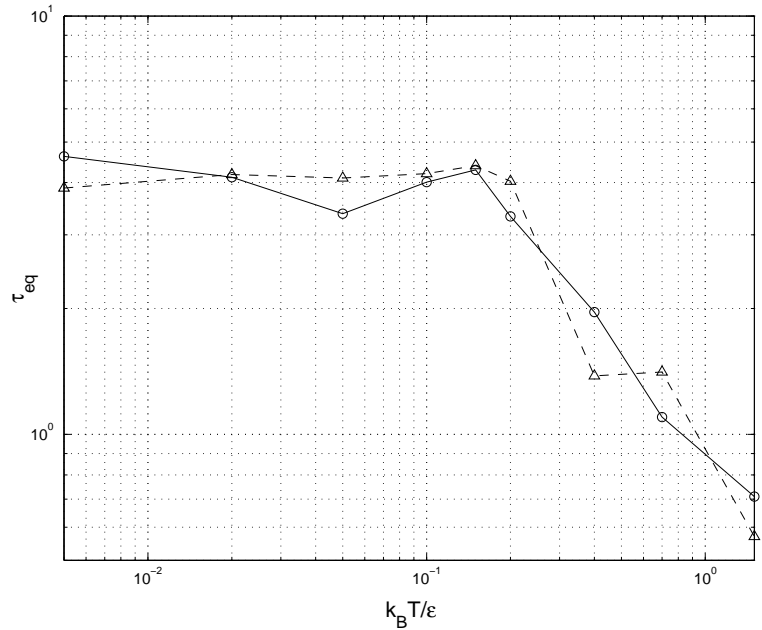}
\caption{Calculated temperature dependence of the Compatible Monte Carlo 
energy relaxation time. The particle number density $\rho\cdot\sigma=0.2$.  
$\circ$ - 16 particles in the periodicity cell 
( $L=80\cdot\sigma$, $M= 32$); $\triangle$ - 256 
particles in the periodicity cell ($L=1280\cdot\sigma$, $M= 512$). The 
size of a subdomain was chosen to be $h_{1}=2.5\cdot
\sigma$.}
\label{f1}
\end{figure}

The temperature dependence of the relaxation time estimated in accordance 
with (\ref{tt3}) is shown in Fig.\ref{f1}. In contrast to the ordinary Monte 
Carlo process the relaxation of CMC is not sensitive to the size of the 
periodicity domain and a fast equilibration implies that coarsening the 
particle number satisfies (in the case of the interaction potential 
(\ref{lj2}) and, hence, in the high temperature limit (\ref{tl4}) the 
requirements for the quality of the coarse variables set. 

In the special case of the one-dimensional fluid, the main contribution 
to the distribution of the particle number $n_{i}^{k}$ at any lattice point 
is expected to come from its nearest neighbors. Therefore a simple 
possible form of the conditional probability table is given by: 
\beq
P_{k}( n_{i}^{k},n_{j}^{k}\mid s^{1}_{ij},\ldots ,s^{l}_{ij})=
P_{k}( n_{i}^{k},n_{i+1}^{k}\mid n_{i-1}^{k},n_{i+2}^{k})
\eeq{eq310} 
Quantities of $n_{i}^{k}$ were calculated at each lattice point $i$ for each 
produced configuration at the next finer level, and from them the probability 
distributions (\ref{eq310}) were accumulated, forming the CP table. Then, 
multilevel Monte Carlo simulation was performed using this CP tables.

A suitable quantity for  comparing  simulation results at different levels 
is the fluctuation $\nu_{k}$ of the particle number in the subdomain of size 
$V^{k}$ (corresponding to the gridpoint of the lattice with meshsize 
$h_{k}=h_{1}\cdot2^{k-1}$):
\beq
\nu_{k}=\frac{<n^{k^{2}}>-<n^{k}>^{2}}{<n^{k}>}
\eeq{eq311}
where the mean values are defined by probabilities $p_{k}(n)$ to display 
the particle number $n^{k}$ in a volume of size $V^{k}$: $<n^{k}>=
\sum\limits_{n^{k}}n^{k}\cdot p_{k}(n^{k})$, $<n^{k^{2}}>=
\sum\limits_{n^{k}}n^{k^{2}}\cdot p_{k}(n^{k})$. 

\begin{figure}[h]
\centering
\includegraphics[width=4in]{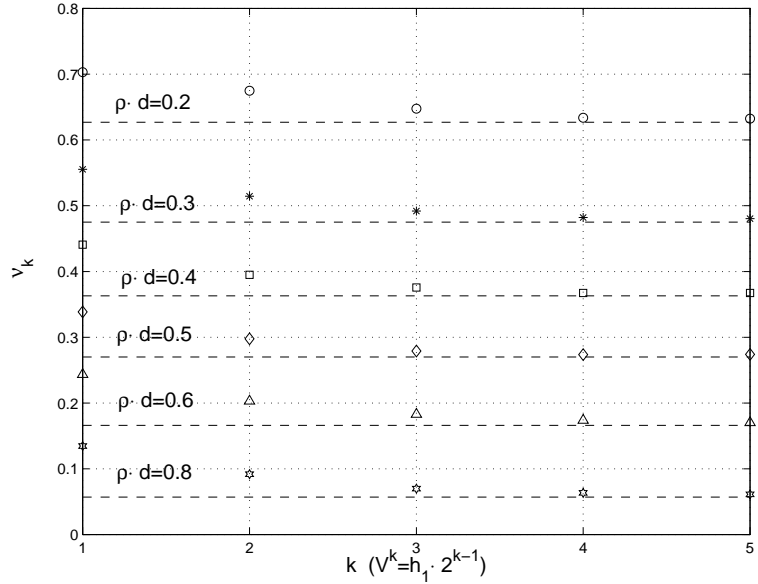}
\caption{Particle number fluctuation at different levels (hard rods fluid, 
$h_{1}=2.5\cdot d$). 
Dashed lines correspond to exact bulk values 
(\ref{tl61}).}
\label{f2}
\end{figure}

The fluctuation of the particle number does not depend on the volume if it 
is sufficiently large (as compared with the correlation length) and is 
related with the isothermal compressibility 
$\kappa=\biggl (-\frac{1}{V}\frac{\partial V}
{\partial P} \biggl )_{T={\it const}}$ \cite{hill}. For 1D hard core
system the equation of state is found analytically (it is known as the Tonk's 
equation  \cite{tonks}) and the isothermal compressibility is given by:
\beq
\nu_{\infty}=\rho\cdot k_{B}T\cdot\kappa=(1-\rho\cdot d)^{2}
\eeq{tl61}

Results of particle number fluctuation multilevel measurements in volumes of 
different sizes for hard rods systems are presented in Fig.\ref{f2}. The 
dependence of this quantity on the subdomain size is explained by the size 
effects and coincides with the expression of the grand canonical 
ensemble \cite{binder1}:
\beq
\nu_{k}=\nu_{\infty}+\frac{C}{V^{k}}
\eeq{eq315}
where $\nu_{\infty}$ is the bulk value of the particle number fluctuation  
(\ref{tl61}). The coefficient $C$ as well as $\nu_{\infty}$ depend on the 
particle number density. One can see from  Fig.\ref{f2} that at high 
temperatures (the interaction model (\ref{tl4}) five levels are enough in 
order to achieve the bulk behavior in subdomains connected with gridpoints 
of the coarsest level.

\begin{figure}[h]
\centering
\includegraphics[width=4in]{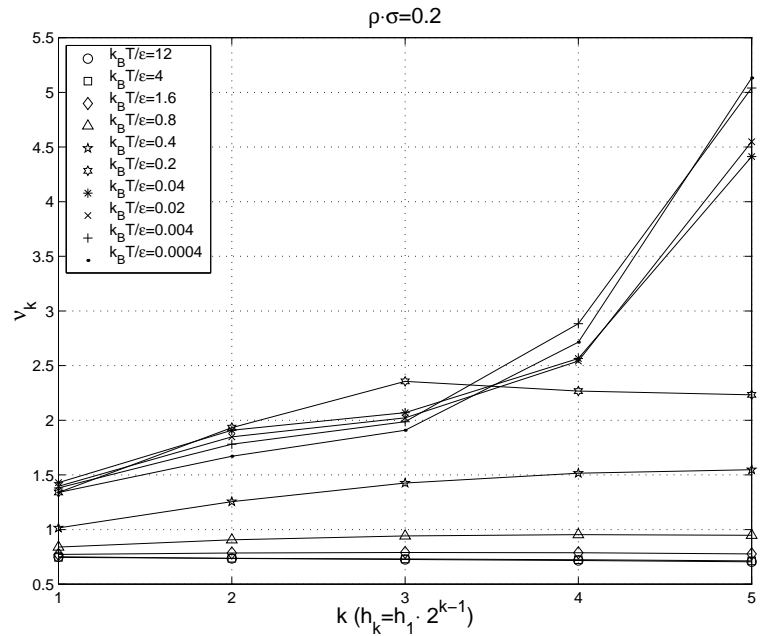}
\caption{Particle number fluctuation at different levels (Lennard-Jones
fluid, $h_{1}=2.5\cdot\sigma$).}
\label{f3}
\end{figure}

In the case of the Lennard-Jones fluid the particle number fluctuations 
depends also on the temperature. The dependence on the meshsize (the 
subdomain volume) for system with the interaction potential (\ref{lj2}) is 
shown in Fig.\ref{f3} for different temperatures. One can see that at high 
temperatures the properties of the Lennard-Jones fluid are similar to the 
hard rods system (see Fig.\ref{f2}). At low temperatures the behavior of 
fluctuations drastically changes. The results obtained indicate that in 
contrast to the hard rods system the Lennard-Jones system loss homogeneity 
at low temperatures on fine scales. 
\section{CONCLUSION}
It follows from the present results that in the framework of the multilevel
Monte Carlo method a suitable choice of coarsening allows to attain the 
macroscopic behavior. The particle number fluctuation in subdomains of 
intermediate size tends to the exact value on increasingly coarser levels. 
The accuracy of averages (including the particle number fluctuation) 
depends only on the CP tables adequacy, which is achieved by the correct 
selection of the variable set.

The advantage of the multilevel Monte Carlo method consists in fast 
convergence of measured mean values due to the selfconsistent equilibration 
on different levels modes of  wave lengths comparable with the simulation 
domain. The computational work on each level is proportional to the number 
of gridpoints and is independent of the particle number associated with 
the gridpoint. It leads to the high speed of the method as compared with 
conventional algorithms.

The equilibrium on fine levels (and in particular the frequency of the 
appearance of a given particle number in the simulation domain) is defined 
by the canonical ensemble configurations on coarsest level.  The particle 
number in fine level simulation domains is variable and its distribution 
imitates the result of the grand canonical ensemble simulation \cite{BI}. 
The multilevel Monte Carlo method is not concerned with the equilibrium 
in accordance with the value of a chemical potential. It opens new way for 
the development of the one-phase approach to the phase-transition problem 
\cite{mart}.

\section{ACKNOWLEDGMENTS}
The research has been supported by Israel Absorption Ministry, project 
No. 6682, by the U.S. Air Force Office of Scientific Research, contract 
No. F33615-97-D-5405, by the European Office of Aerospace Research and 
Development (EOARD) of the U.S. Air FORCE, contract No. F61775-00-WE067, 
by Israel Science Foundation Grant No. 696/97 and
by the Carl F.Gauss Minerva Center for Scientific Computation at the 
Weizmann Institute of Science. 



\end{document}